\documentclass[journal=jacsat,manuscript=article]{achemso}

\usepackage[version=3]{mhchem} 
\usepackage{color}
\usepackage{amsmath}
\usepackage{latexsym}
\usepackage{amsfonts}
\usepackage{hyperref}
\usepackage{pdfpages}

\newcommand{\ma}{\sigma}
\newcommand{\gm}{\gamma}

\newcommand{\dt}{\delta}

\newcommand{\f}{\frac}
\newcommand{\ta}{\theta}

\newcommand{\A}{\alpha}

\newcommand{\w}[1]{\omega_{#1}}

\newcommand{\alg}[1]{\begin{align}#1\end{align}}

\author{Taekoo Oh}
\affiliation[RIKEN]{RIKEN Center for Emergent Matter Science (CEMS), Wako, Saitama 351-0198, Japan}
\email{taekoo.oh@riken.jp}
\author{Naoto Nagaosa}
\affiliation[RIKEN]{RIKEN Center for Emergent Matter Science (CEMS), Wako, Saitama 351-0198, Japan}
\email{nagaosa@riken.jp}

\title[An \textsf{achemso} demo]
  {Unraveling the dynamics of magnetization in topological insulator-ferromagnet heterostructures via spin-orbit torque}

\abbreviations{IR,NMR,UV}

\begin{document}

\begin{tocentry}

\includegraphics[width=\columnwidth]{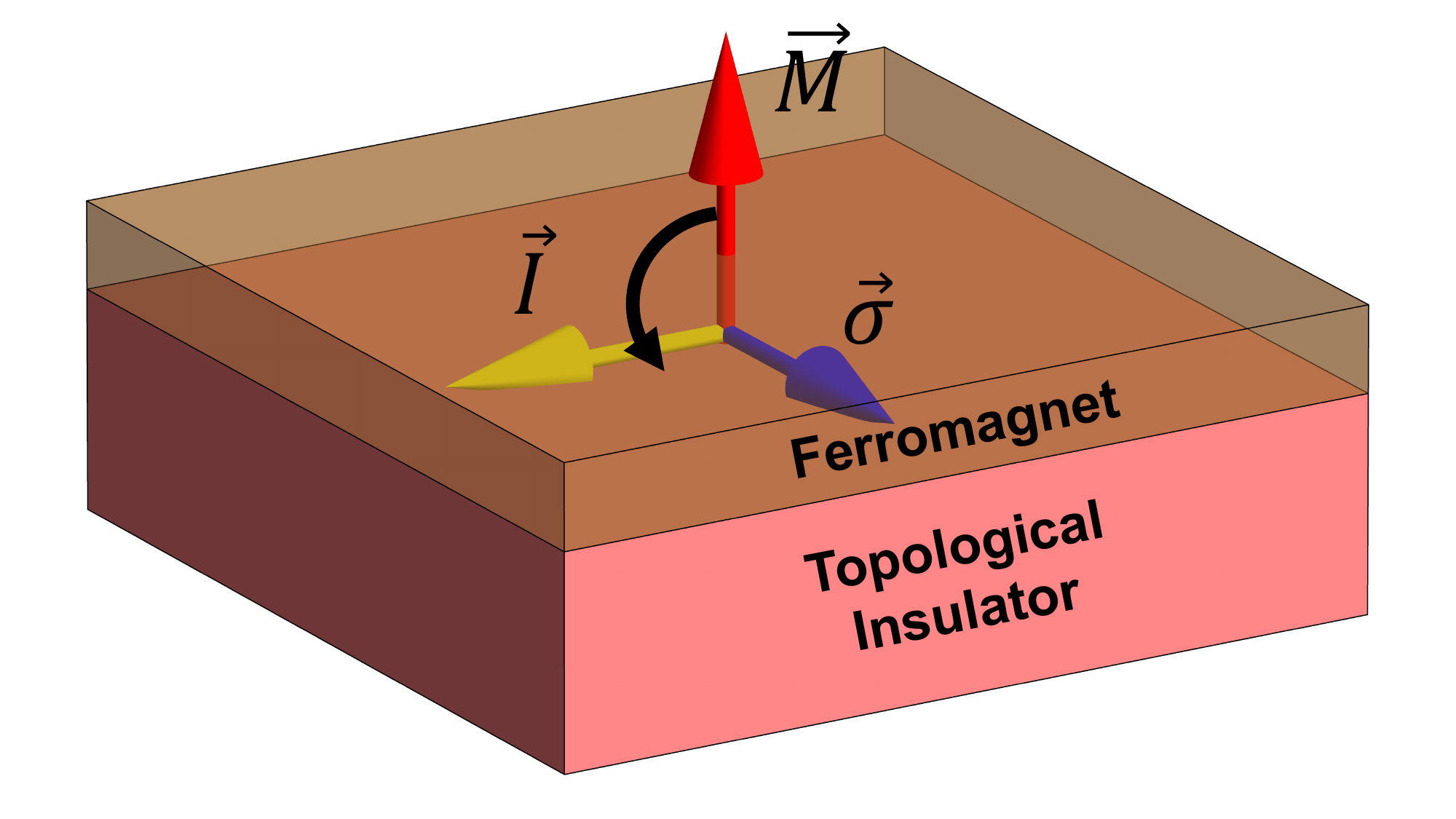}

\end{tocentry}

\begin{abstract}
Spin-orbit coupling stands as a pivotal determinant in the realm of condensed matter physics. 
In recent, its profound influence on spin dynamics opens up a captivating arena with promising applications. 
Notably, the topological insulator-ferromagnet heterostructure has been recognized for inducing spin dynamics through applied current, driven by spin-orbit torque. 
Building upon recent observations revealing spin flip signals within this heterostructure, our study elucidates the conditions governing spin flips by studying the magnetization dynamics. 
We establish that the interplay between spin-anisotropy and spin-orbit torque plays a crucial role in shaping the physics of magnetization dynamics within the heterostructure. 
Furthermore, we categorize various modes of magnetization dynamics, constructing a comprehensive phase diagram across distinct energy scales, damping constants, and applied frequencies. 
This research not only offers insights into controlling spin direction but also charts a new pathway to the practical application of spin-orbit coupled systems.
\end{abstract}

\section{Keywords}

Spintronics, Magnetization dynamics, Topological insulator, Spin-orbit torque.

\section{Introduction}

~~~~Spin-orbit coupling (SOC), acknowledged as one of the fundamental interactions in materials~\cite{witczak2014correlated}, reveals a fertile landscape within condensed matter physics. A standout illustration of its impact is evident in the burgeoning field of spintronics, where SOC-driven spin dynamics unveils a promising horizon for practical applications.
Accordingly, recent attention has been drawn to topological insulator-ferromagnet (TI-FM) heterostructures, as evinced by notable reports~\cite{hasan2010colloquium,qi2011topological,pi2010tilting,liu2011spin,kim2013layer,che2020strongly}. 
The surge in interest stems from their remarkable efficiency in catalyzing spin dynamics, primarily attributed to the spin-orbit torque (SOT) induced by an applied current~\cite{brataas2012current,shao2021roadmap}. 

These heterostructures exhibit intriguing phenomena that closely follow its intricate dynamics of spins.
For instance, an emergent inductance from the motive force in spiral magnets~\cite{nagaosa2019emergent,yokouchi2020emergent,kitaori2021emergent} was extended to the TI-FM heterostructures~\cite{ieda2021intrinsic,yamane2022theory,araki2023emergence}. Nonreciprocal transport phenomena associated with the SOC was explored~\cite{yasuda2016large,yasuda2017current}.The significant increase of Curie temperature was reported as well~\cite{alegria2014large,katmis2016high,wang2020above,ou2023enhanced}. 
Notably, it has been acknowledged that the SOT at the TI-FM interface is robust enough to flip magnetization and induce the sign change of Hall Effect~\cite{fan2014magnetization,mellnik2014spin,han2017room,wang2017room,mogi2021current}. Such properties hold promise for applications in spintronics devices, particularly in utilizing topological SOTs for magnetic memories~\cite{yokoyama2010theoretical,garate2010inverse,pesin2012spintronics,araki2021intrinsic,wu2021magnetic,yamanouchi2022observation}..

Motivated by such intriguing phenomena due to the SOC and SOT, this study explores magnetization dynamics in TI-FM heterostructures with an applied current, aiming to unveil the conditions for magnetization flip. 
Considering both spin-anisotropy and SOT, we establish the equilibrium state of magnetization and develop a model describing its dynamics under direct current (DC) or alternating current (AC). 
In the absence of damping, we identify oscillating, faltering, and flipping modes for DC, with the latest inducing magnetization flip. 
The choice between modes is determined by the relative strengths of SOT to spin-anisotropy. 
With the introduction of damping, the occurrence of spin flip hinges on the duration elapsed until the crossover from flipping to oscillating modes takes place.

For AC, on the other hand, we discern three final states---adiabatic, resonating, and chaotic. 
We classify five distinct modes in the adiabatic state at low frequencies, which is derived from DC modes.
By providing the phase diagrams of adiabatic modes, we illustrate that the modes are determined by SOT, spin-anisotropy, driving frequency, and initial driving phase.
Lastly, we explore their transition to resonating and chaotic states at higher frequencies in the viewpoint of the Fourier transform, revealing that the periodic array of peaks gives rise to the chaotic state. 
By overhauling the dynamics, we provide insights into the complex dynamics of magnetization in TI-FM heterostructures.

\begin{figure}[t]
	\centering
	\includegraphics[width=\columnwidth]{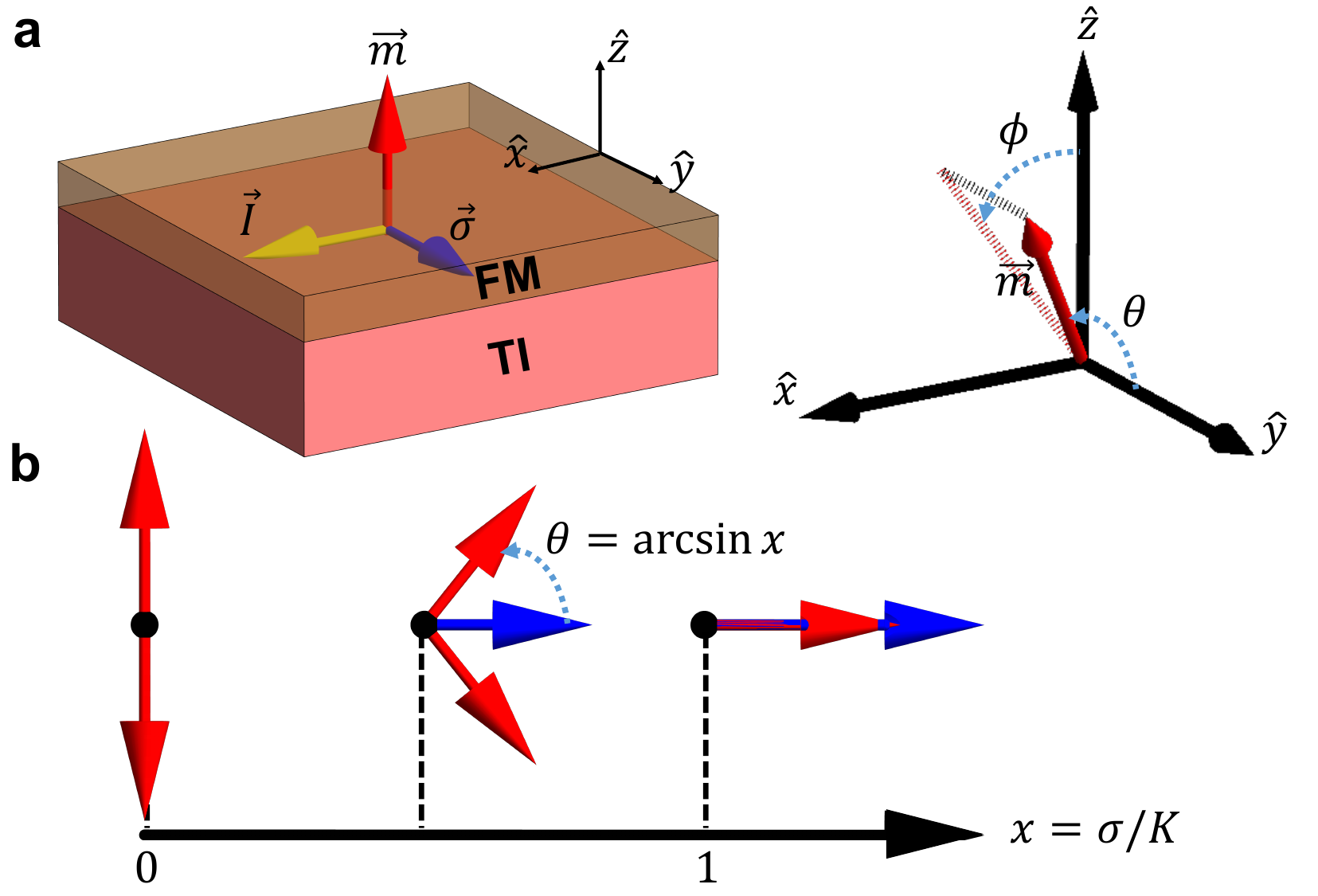}
	\caption{{\bf Magnetization dynamics of TI-FM heterostructure. }(a) The schematics of the TI-FM heterostructure. (b) The equilibrium state of magnetization in the different relative strengths of SOT $x=\ma/K$. }
	\label{fig:1}
\end{figure}

\section{Results}

\subsection{Model and its equilibrium}

The physical configuration is illustrated in Fig.~\ref{fig:1}(a), where a current flows along the $x$-axis, the itinerant electron spin aligns with the $y$-axis, and the ferromagnet's magnetization lie along the $+z$-axis at $t=0$. The polar and azimuthal angles with respect to the $y$-axis are denoted as $\ta$ and $\phi$. We represent the magnetization as $\vec{m} = m(\sin\ta\sin\phi,\cos\ta,\sin\ta\cos\phi)$ and the itinerant electron spin as $\vec{\ma} = \ma(t)(0,1,0)$.
For simplicity, we exclusively focus on the dynamics of the single $\vec{m}$, and ignore that of $\vec{\ma}$. 
This is justified by the stronger spin-momentum locking of the TI surface state compared to anisotropy~\cite{araki2023emergence,daalderop1990first,konig2008quantum,cullity2011introduction,qi2011topological,xu2020electronic}. 
For instance, Ni has the easy-axis anisotropy energy about $2.7~\mu$eV$/$atom in the bulk~\cite{fritsche1987relativistic} and $4.0~$meV$/$atom in the monolayer~\cite{gay1986spin}, while the spin-orbit coupling of Bi$_2$Se$_3$ is estimated as $\sim 1.2~$eV$/$atom. Accordingly, we set our unit of energy to be $1.0~ \mu$eV $\sim 2.4~$GHz and the unit of time to be $(2.4~$GHz$)^{-1} \approx 0.42~$ns.
The spin flip is then denoted as the switching of spin direction between $+z$ and $-z$ half space.

Two crucial potential energies emerge: $V_A = -\frac{K}{2}m_z^2$ $(K>0)$, representing magnetic anisotropy in the ferromagnet, and $V_S = -\gamma \vec{m}\cdot\vec{\ma}$  signifying the Rashba-field-like~\cite{liu2012current,manchon2019current} (or domain-damping-like)~\cite{kim2019skyrmions} SOT. 
The equilibrium state of magnetization is obtained, showcased in Fig.~\ref{fig:1}(b) for fixed $\vec{\ma}$. 
Normalizing $m = \gm = 1$, key energy scales become the anisotropy energy $K$ and SOT $\sigma$. The dimensionless parameter $x=\frac{\sigma}{K}$ is defined. As $x$ increases from 0, the equilibrium states initially align along $\pm z$-axis and gradually rotate towards the $y$-axis. In terms of angles, $\sin\ta = x, \phi = \pm \pi/2$. Before reaching the $y$-axis, the equilibrium states are twofold degenerate. For $x\geq1$, the degenerate equilibrium states converge at the $y$-axis.

To describe magnetization dynamics out of equilibrium, the Lagrangian density is considered:
\alg{
	\mathcal{L} = \mathcal{L}_B - V_A - V_S, ~~
	\mathcal{L}_B = m(1-\cos\ta)\dot\phi.
}
Note that $\mathcal{L}_B$ is the Berry phase term, making polar and azimuthal angles conjugate~\cite{auerbach2012interacting,tatara2014phasons}. Thus, as magnetization rotates from the $z$-axis to its equilibrium state, deviations from the $yz$-plane induce complex motion. Introducing Gilbert damping through the Rayleigh dissipation function~\cite{gilbert2004phenomenological}, $\mathcal{R} = \frac{\eta}{2}\f{\partial\vec{m}}{\partial t}\cdot\f{\partial\vec{m}}{\partial t}$, with the dimensionless damping constant $\eta$, yields the equations of motion:
\alg{
	(1+\eta^2)\dot\ta =& -
	\f{K}{2}\sin\ta \sin2\phi +\f{\eta K }{2}\sin2\ta\cos^2\phi  - \eta \sigma(t) \sin\ta,\text{ and}\\
	(1+\eta^2)\dot\phi =& - K \cos\ta\cos^2\phi + \sigma(t)  - \f{\eta K}{2}\sin2\phi.
}
The initial conditions are set at $\ta=\pi/2$ and $\phi=0$. This system of equations governs physics for both DC and AC scenarios. The numerical computation is performed by Mathematica.

It should be noted that the field-like SOT $\sigma(t)$ precesses the magnetization around $y$-axis while the damping-like SOT $\eta\sigma(t)$ cants the magnetization toward $y$-axis. This is consistent with the Landau-Lifshitz-Gilbert equation. 
It is noteworthy that the relation between the damping constant $\eta$ and the Gilbert damping constant $\A$ is $\alpha = \eta \gamma$, where $\gamma$ is the gyromagnetic ratio~\cite{gilbert2004phenomenological}. We already set $\gamma = 1$, resulting in $\alpha = \eta$. As a consequence, the field-like SOT is responsible for the spin flip, whereas the damping-like SOT deters it. 

\subsection{Magnetization dynamics under DC}

In the exploration of magnetization dynamics under DC, we initiate with a straightforward scenario where $\alpha=0$ and $\sigma(t) = 0$ for $t<0$ and $\sigma(t)=\sigma$ for $t\geq 0$. Lacking damping, we anticipate permanent magnetization precession, described by the simplified equations:
\alg{
	\dot\ta = - K \sin\ta \sin\phi\cos\phi,~~ \dot\phi = - K \cos\ta\cos^2\phi + \sigma.
}
These equations are intricate to solve analytically, but insight can be gained by examining two limiting cases, $x \ll 1$ and $x \gg 1$ with $\alpha=0$. In the first limit, we find oscillatory behavior:
\alg{
	\ta = \f{\pi}{2} + x(\cos(Kt)-1), ~~
	\phi = x\sin(Kt).
}
Both $\ta$ and $\phi$ oscillate with amplitude $x$ and frequency $K$, termed the {\it oscillating mode}. In the second limit:
\alg{
	\ta = \f{\pi}{2} + \f{1}{4x}(\cos(2\sigma t)-1),~~
	\phi =  \sigma t.
}
Here, $\ta$ oscillates with amplitude $1/(4x)$ and frequency $2\sigma$, while $\phi$ monotonically increases. This implies the spin precession about the $y$-axis and a spin flip occurring at  period of $\pi/\sigma$, termed the {\it flipping mode}. Comparing the frequency of $\ta$ dynamics in each mode, we speculate that the transition from the oscillating to the flipping modes might occur at $x=1/2$ where $K=2\sigma$.

\begin{figure}[ht]
	\centering
	\includegraphics[width=0.9\columnwidth]{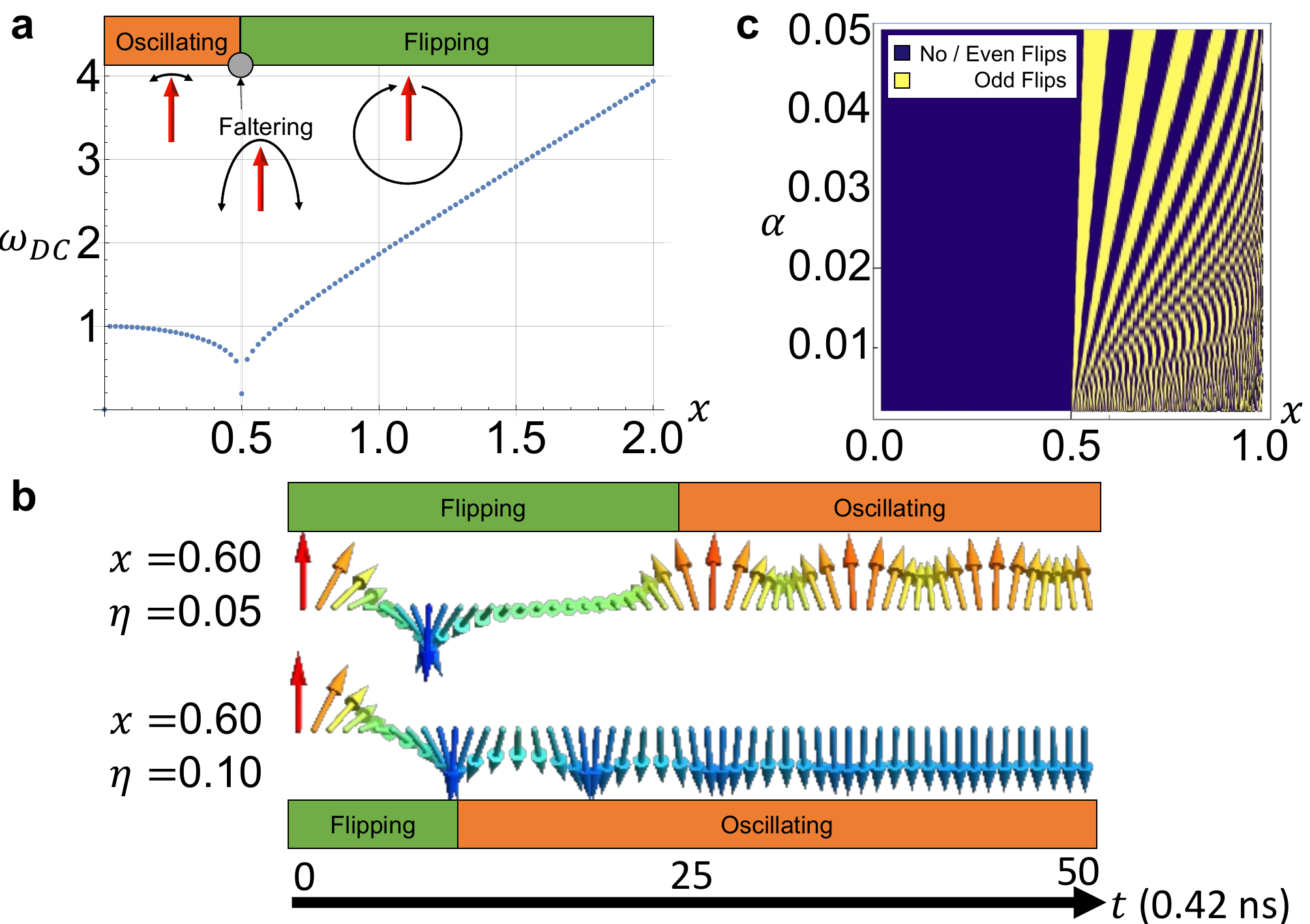}
	\caption{{\bf The magnetization dynamics under DC. }(a) Computed frequencies of oscillating $(x<0.5)$, faltering $(x=0.5)$, and flipping modes $(x>0.5)$ under DC without damping in varying $x=\sigma/K$. (b) Dynamics of magnetization at $x=0.6,\alpha=0.05$ (lower) and $x=0.6, \alpha=0.1$ (upper) in time. Chromatically, the red indicates $+z$, the green indicates in-plane, and the blue indicates $-z$ directions of magnetization. The direction inward the paper is the $y$ direction. (c) A 2D phase diagram of the final states in varying $x$ and $\alpha$. The dark blue indicates the even-flipped final  state, and the yellow indicates the odd-flipped final state. }
	\label{fig:2}
\end{figure}

The guess can be verified by numerical computations of frequency in units of $K$ for $\sigma(t) = \sigma$ for $t\geq 0$ and $\alpha=0$ in Fig.~\ref{fig:2}(a). The frequency variation exhibits $\omega_{DC} \approx K = 1$ for $x\ll 1$ and $\omega_{DC} \approx 2\sigma$ for $x\gg 1$. 
The singularity at $x=1/2$ indicates the expected transition. 
Only at this transition point, the spins are faltering between $z$ axis and $xy$-plane in period, so we call this {\it faltering mode.}
For DC, we focus on the oscillating and flipping modes, as the faltering mode requires a fine tuning. 

When damping $\alpha$ is introduced under DC, the magnetization eventually attains an equilibrium state.
For $x\geq 1$, the equilbrium state aligns with $y$-axis, so the spin flip is absent.
However, since two equilibrium states exists for $x<1$, the selection between them occurs, which can flip the spin. 
The selection is determined by both $x$ and $\alpha$.
Figure~\ref{fig:2}(b) displays two examples of dynamics leading to an even-flipped ($x=0.6, \alpha=0.05$) and odd-flipped ($x=0.6, \alpha =0.1$) final state, respectively. 
The magnetization in the even-flipped state stays in $+z$ half space, while that in the odd-flipped state stays in $-z$ half space.
In both examples, owing to the damping effect, the crossover from the flipping to the oscillating mode occurs after a specific time $\tau_c$.
In Fig.~\ref{fig:2}(b), for the former, after $\tau_c \approx 10.8$ ns, while for the latter, $\tau_c \approx 4.2$ ns. 
Empirically, we find that $\tau_c \propto e^{4.53x} \alpha^{-1}$ for small $\alpha$. [See Supporting Information (SI).]
These show that the variation in $\tau_c$ caused by the interplay of $x$ and $\alpha$ serves as a determining factor for the final state. 

We further delve into the relation of the final state with $x$ and $\alpha$ by a 2D phase diagram in Fig.~\ref{fig:2}(c). 
When $x<1/2$, exclusively the even-flipped final state manifests, whereas for $x\geq 1/2$, both odd-flipped and even-flipped states emerge, forming a fan-like configuration. 
It should be noted that the fan-like configuration is not clear in $\alpha<0.01$ due to the resolution.
The exclusive appearance of even-flipped state for $x<1/2$ is due to the absence of spin flip in the oscillating mode.
However, in the case of $x\geq 1/2$, the spin flip can occur multiple times before the crossover from the flipping to the oscillating modes takes place.
If the crossover time exceeds half of the flipping mode period, the spin flips to $-z$ half space.
With a longer crossover time surpassing a full flipping mode period, the spin flips twice, returning to $+z$ half space.
By extending the crossover time further, the spin flips repeatedly. 
Even (odd) numbers of spin flips lead to an even-flipped (odd-flipped) final state, giving rise to the distinctive fan-like feature in $x\geq 1/2$.
This pattern persists until $x=1$, where the equilibrium state converges to $y$-axis.

\subsection{Magnetization dynamics under AC}

\begin{figure}
	\centering
	\includegraphics[width=0.7\columnwidth]{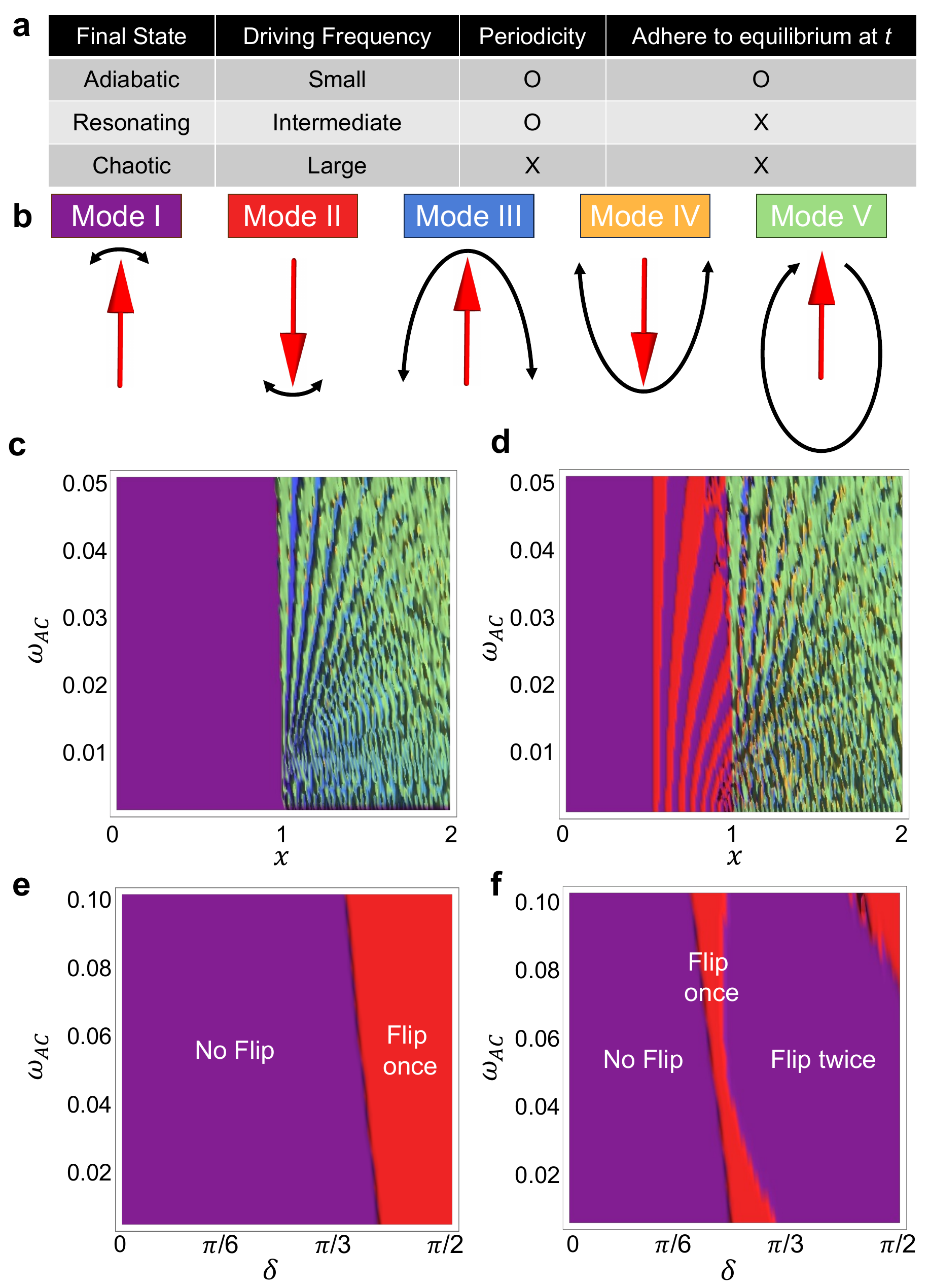}
	\caption{{\bf The modes in the adiabatic state under AC.} (a) The classification and comparison of final states. The final states transit with the driving frequency of AC, from adiabatic to resonating, and to chaotic state in sequence. While adiabatic and resonating states have periodicity, chaotic state does not. While adiabatic state adheres to the equilibrium state at each $t$, others do not. (b) Schematics of distinct AC adiabatic modes after decay. (c-d) 2D phase diagrams of adiabatic modes in $x$ and $\w{\text{AC}}$ with $\alpha=0.03$, at (c) $\dt=0$ and (d) $\dt=\pi/2$. (e-f) 2D phase diagrams of adiabatic modes in $\dt$ and $\w{\text{AC}}$ at (e) $x=0.6$ and (f) $x=0.8$. The violet denotes Mode I, the red denotes Mode II, the blue denotes Mode III, the yellow denotes Mode IV, and the green denotes Mode V.}
	\label{fig:3}
\end{figure}

Based on above results, we here explore the magnetization dynamics under AC, which is given by $\sigma(t) = \sigma \sin(\omega_{\text{AC}}t+\delta)$ for $t\geq 0$ and $\sigma(t) = 0$ for $t<0$. $\w{\text{AC}}$ is the driving frequency, and $\delta$ is the initial phase of AC.
The initial state of the magnetization is again aligned with $+z$-axis.
Owing to the damping $\alpha$, the relative strength of SOT $x$, and driving frequency $\w{\text{AC}}$, the initial state overcomes the irregular dynamics and transits to the final states after some time.
We identify three distinct final states in Fig.~\ref{fig:3}(a): adiabatic, resonating, and chaotic states.
Both adiabatic and resonating states are the steady states coming after the decay of initial state. 
At each time $t$, $\ma(t)$ determines an equilbrium state as shown in Fig.~\ref{fig:1}(b).
The adiabatic state denotes that the magnetization mostly adheres to the equilibrium state at each time.
The resonating state denotes that the magnetization dynamics is periodic but detach from the equilibrium state at each time.
On the other hand, the chaotic state is evolved from the initial state, retaining its irregularity, in which the magnetization never has a periodic motion.

We primarily observe the adiabatic states for low-frequency or strong damping regime. 
We observe five distinct modes in the adiabatic states: I) an even-flipped oscillating mode, II) an odd-flipped oscillating mode, III) an even-flipped faltering mode, IV) an odd-flipped faltering mode, and V) a periodically flipping mode. 
These modes are depicted in Fig.~\ref{fig:3}(b). The even-flipped (odd-flipped) oscillating mode or Mode I (II) is the oscillation of magnetization within $+z$ ($-z$) half space. 
The even-flipped (odd-flipped) faltering mode or Mode III (IV) is the repeated faltering of magnetization between $xy$-plane and $+z$ ($-z$) half space. 
The periodically flipping mode or Mode V is the repeated magnetization flip.
Notably, Modes I and II (or III and IV) are almost identical, but differ in the number of spin flips before decaying to the adiabatic state. Only Mode V shows the continuous spin flip in time in its adiabatic state. As one can expect from their names, the adiabatic modes originate from DC modes. Modes I and II come from oscillating mode, Modes III and IV come from the faltering mode, and Mode V comes from the flipping mode. 

The modes are chosen by the interplay of  $x$, $\w{\text{AC}}$, $\dt$, and $\alpha$. Both
$x$ and $\w{\text{AC}}$ underscore their importance after decaying to adiabatic states while $\dt$ and $\alpha$ play a pivotal role during the decay process.
Primary investigation is performed by phase diagrams for the modes in $x$ and $\omega_{\text{AC}}$ at $\alpha=0.03$ with different $\dt$ presented in Figs.~\ref{fig:3}(c,d). 
For $x<1$, Modes I and II appear, while for $x\geq 1$, Modes III, IV, and V manifest. This transition occurs around $x\sim 1$ since $x=1$ is where two equilibrium states meet at the $y$-axis. 
Specifically, for $x<1$, as the equilibrium position does not reach $y$-axis at any time, the magnetization oscillates within either $+z$ or $-z$ half space. 
For $x\geq 1$, however, as the equilibrium reaches $y$-axis, the magnetization either falters on the $xy$-plane or flips its position between $+z$ and $-z$ half spaces periodically. 

$\omega_{\text{AC}}$ also plays a role in determining the modes. 
Primarily, $\omega_{\text{AC}}$ chooses the faltering and flipping modes when $x\geq 1$, since it determines the time duration $\tau_y$ that the equilibrium state at each time stays at the $y$-axis. 
One can obtain the duration by finding the maximum $\tau_y$ satisfying $\ma(t)\geq 1$ in $t\in[t_0,t_0+\tau_y]$.
In the case of sinusoidal $\ma(t)$, this can be expressed as $\tau_{y} = \f{1}{\w{\text{AC}}}(\pi - 2\zeta)$, where $\zeta = \arcsin(1/x) \in (0,\pi/2]$.
For $\tau_y$, as the equilibrium state is at the $y$-axis, the magnetization modulates near the $y$-axis. 
After $\tau_{y}$, the equilibrium position is divided again and deviates away from the $y$-axis. 
Then, depending on its modulated position, the magnetization chooses one of the equilibrium state, which leads to either faltering or flipping modes.
Additionally, the window of $x$ for every mode at higher $\w{\text{AC}}$ is opened up wider than that at lower $\w{\text{AC}}$, showing the fan-like feature in Figs.~\ref{fig:3}(c,d). 
Unlike the faltering mode under DC, the window of Modes III and IV, originating from the faltering mode, expands to the finite range of $x$.
Lastly, it is noteworthy that the phase of $\w{AC}=0$ in Figure~\ref{fig:3}(d) is the same as the phase in $\alpha=0.03$ line of Fig.~\ref{fig:2}(c). This substantiates that adiabatic modes can be derived from DC modes.

On the other hand, $\dt$ acts as a switch to turn on the spin flip during the decay process to adiabatic states.
Specifically, comparing Figs.~3(b) to (c), Modes II and IV barely appear when $\dt=0$, which means that the spin flip is turned off near $\dt=0$. This happens because the average of $\sigma(t)$ during the decay time is small, so the spin cannot flip before decaying the adiabatic state.
This can be supported by Figs.~\ref{fig:3}(e-f). 
In Fig.~\ref{fig:3}(e), we present phase diagrams for the modes in $\delta$ and $\omega_{\text{AC}}$ at $x=0.6$. 
The transition from Modes I to II occurs once during the increase of $\dt$. This happens because the spin flip is turned on by finite $\dt$.
Increasing $x$, the spin flips more times just as in DC case, so the repeated transition between Modes I and II can also be observed in Fig.~\ref{fig:3}(f). 
We should note that the increase of $\alpha$ reduces the decay time.
Above arguments holds also for the triangular wave instead of sinusoidal wave. [See SI.]

\begin{figure}[t]
	\centering
	\includegraphics[width=\columnwidth]{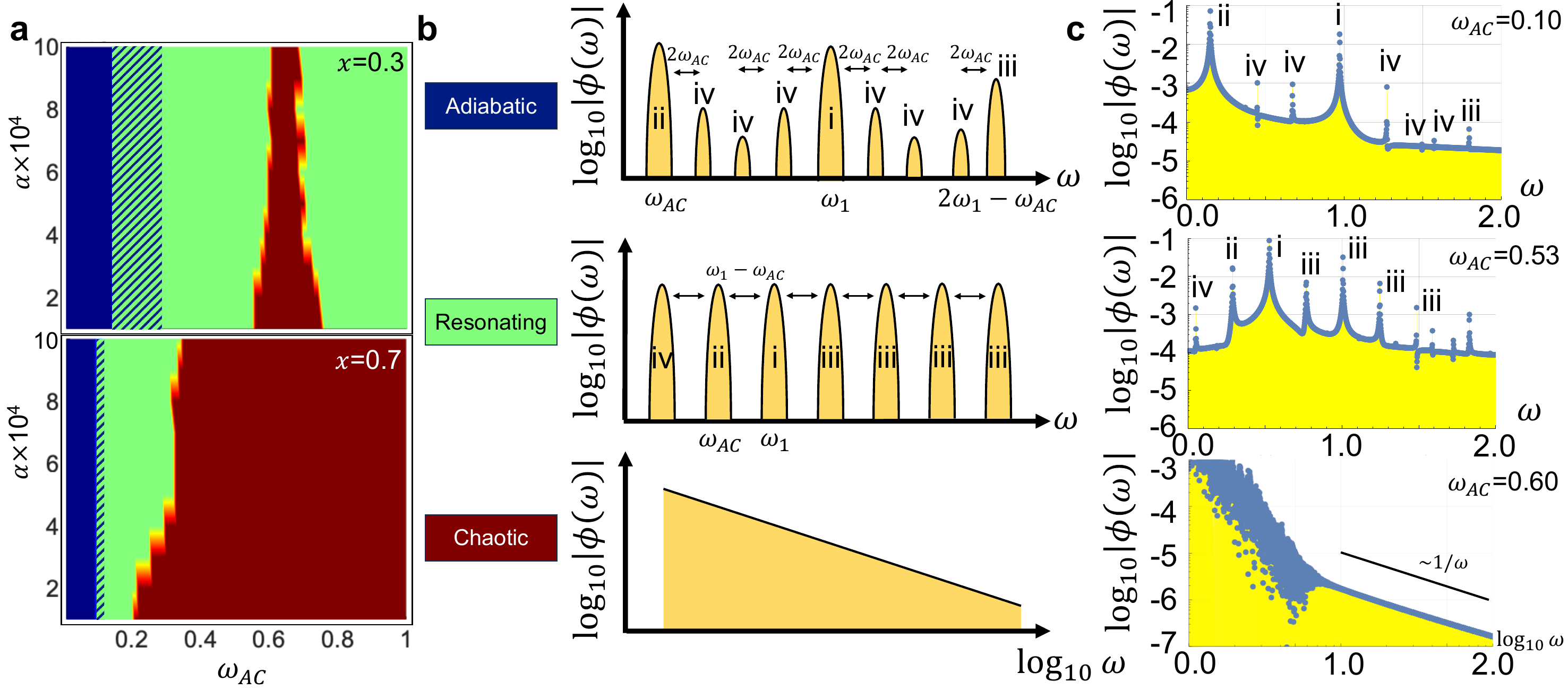}
	\caption{{\bf The transition from adiabatic to resonating and chaotic states under AC.} (a) 2D phase diagrams of the final states in $\w{\text{AC}}$ and $\alpha$. The adiabatic state is in dark blue, the resonating state is in green, and the  chaotic state is in dark red. The top panel is at $x=0.3$ and the bottom panel is at $x=0.7$. (b) The schematics of change in Fourier transform in for each state. (c) The corresponding examples to each state of (b) at $x=0.3$ and $\alpha=0$.}
	\label{fig:4}
\end{figure}

We move on to high-$\omega_{\text{AC}}$ and low-$\alpha$ regime to investigate ferromagnetic magnetoresonance (FMR).
One could expect that the resonance frequency is closely related to the frequencies $\omega_{\text{DC}}$ in Fig.~\ref{fig:2}(a).
In fact, when $\omega_{\text{AC}}$ approaches  $\omega_{\text{DC}}$, the oscillating amplitude becomes larger, and eventually the adiabatic state changes to the resonating state and to the chaotic state sequentially.
By checking the final state after $t \sim 1~$ms, we present 2D phase diagrams in $\w{\text{AC}}$ and $\alpha$ at $x=0.3$ and $0.7$ in Fig.~\ref{fig:4}(a). Both diagrams show similar behavior. It begins with the adiabatic state at low frequencies, confronts the transition to the resonating state, and reaches the chaotic state at a certain range of $\w{\text{AC}}$. For $x=0.7$, the window of chaotic states is opened up widely as SOT becomes large.

We further illuminate the transition of states by Fourier transform. 
The transition is illustrated in Fig.~\ref{fig:4}(b), where the change in $|\phi(\omega)|$ by $\omega_{\text{AC}}$ is schematically presented. Here, $\phi(\omega)$ corresponds to the Fourier transform of azimuthal angle $\phi(t)$ from $t=0$ to $t=0.21 ~ $ms ($5\times10^5$ unit times).
Figure~\ref{fig:4}(c) showcases representative examples of Fourier transforms at $x=0.3$ and $\alpha=0$. 
For adiabatic and resonating states, four distinct peaks are observed in the top panels: i) a main peak at $\omega_{1}$ derived from $\omega_{\text{DC}}$, ii) another main peak driven by AC at $\omega_{2}=\omega_{\text{AC}}$, iii) induced peaks from i) and ii) at $\omega_{3}= \omega_{\text{1}} + n(\omega_{\text{1}}-\omega_{\text{AC}})$, and iv) subpeaks at $\omega_{4} = |\omega_{1,2,3} \pm 2n\omega_{\text{AC}}|$ ($n \in \mathbb{N}$). 
Notably, as the time interval is elongated from 0.21 ms, Peaks i and iii decay while Peaks ii and iv gain intensity. [See SI.] 
This means that Peaks i and iii are related to the decay process while Peaks ii and iv are related to the stable state after decaying.
Increasing $\w{\text{AC}}$ to the higher frequency, the change only occurs to the distance between peaks, as the frequency of Peak i decreases and that of Peak ii increases.
Thus, adiabatic and resonating states are indistinguishable solely by Fourier transform.
This is substantiated by the phase diagrams in Fig.~\ref{fig:4}(a), where the transition from adiabatic and resonating states is not distinctly delineated.
This behavior is consistent for all modes in Fig.~\ref{fig:3}(b).
Near the chaotic state, the peaks form array in a period of $\w{\text{AC}}-\w{1}$ as shown in the middle panels.
As $\omega_{\text{AC}}$ increases more, chaos sets in, destroying all peaks and being $\phi(\omega)\propto 1/\omega$ as shown in the bottom panels. 
This behavior does not change although the range of time is elongated.

\section{Discussion}

We address here about the typical values of parameters in the realistic systems. The typical value of ferromagnetic anisotropy is $1-10~\mu$eV/atom~\cite{mogi2021current}, that of the current density in the experiments is $\sim 10^7~$A$\cdot$cm$^{-2}$, and that of the itinerant spin polarization by Rashba-Edelstein Effect is estimated about $10^{-4} \hbar $ per unit cell~\cite{chen2019edelstein}. The typical value of Gilbert damping constant is $\sim 10^{-3} - 10^{-2}$~\cite{barati2013calculation}, and that of the unit of AC frequency is estimated about $1 - 10$ GHz.

So far, we discuss the magnetization dynamics at the interface of TI-FM heterostructure. Under DC, we observe the oscillating and flipping modes without damping, which is determined by the relative strength of anisotropy and SOT.
With damping, the number of spin flips is determined by the duration of the crossover from flipping to the oscillating mode.
Under AC, we observe five distinct modes in the low frequency regime and their evolution to the resonating and chaotic states in the high frequency regime. 
Although we mainly discuss the TI-FM heterostructure due to its high efficiency, our work can be applied to the general systems with strong Rashba spin-orbit coupling. 
This is because the assumption underlying in our work is only that the current carries finite spin due to the Rashba-Edelstein Effect. Our work offers insights into the spin control by spin-orbit coupling, underscoring the practical aspects of the world of spintronics.

\begin{acknowledgement}

We thank Wataru Koshibae for the fruitful discussions. This work was supported by JST, CREST Grant Number JPMJCR1874, Japan.

\end{acknowledgement}

\begin{suppinfo}

This material is available free of charge via the Internet at https://pubs.acs.org/

{\indent $\bullet$ The dependence of crossover time from flipping to oscillating mode on $x$ and $\alpha$, the dependence of decay time to adiabatic modes on $\alpha$, the comparison between triangular and sinusoidal waves, and the time evolution of peaks in Fourier transform.}

\end{suppinfo}



\section*{Supplementary material}

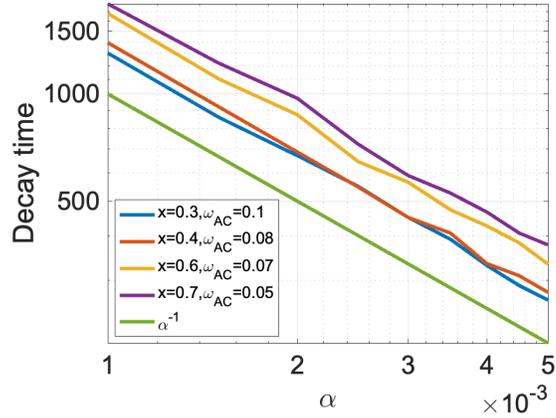

Figure 2: **The decay time in $\alpha$.** The decay time in $\alpha$ with different $x$ and $\omega_{\text{AC}}$.

# The comparison of triangular and sinusoidal waves

We mainly discuss the magnetization dynamics under sinusoidal waves of AC. One could replace the sinusoidal waves with triangular waves. The triangular wave is given by

$$\sigma(t) = |\text{mod}[\frac{2\omega_{\text{AC}}\sigma}{\pi}(t-t_0), 4\sigma] - 2\sigma| - \sigma. \tag{1}$$

The amplitude of this function is $\sigma$, and the frequency of this function is $\omega_{\text{AC}}$. At $t_0 = \pi/(2\omega_{\text{AC}})$, the function becomes sine-like, while at $t_0 = 0$, the function becomes cosine-like, as shown in Fig. 3(a).

We set $\omega_{\text{AC}} = 2\pi/240 \approx 0.026$, $\alpha = 0.03$, and compare the adiabatic modes for sinusoidal and triangular waves in Fig. 3(b). In the manuscript, we describe that Modes II and IV does not appear under sine waves. The argument is also consistent with triangular waves, since the phase diagram under sine-like triangular waves does not show Modes II and IV as well. The difference between sinusoidal and triangular waves is that the window for each mode widens up for triangular waves.

3

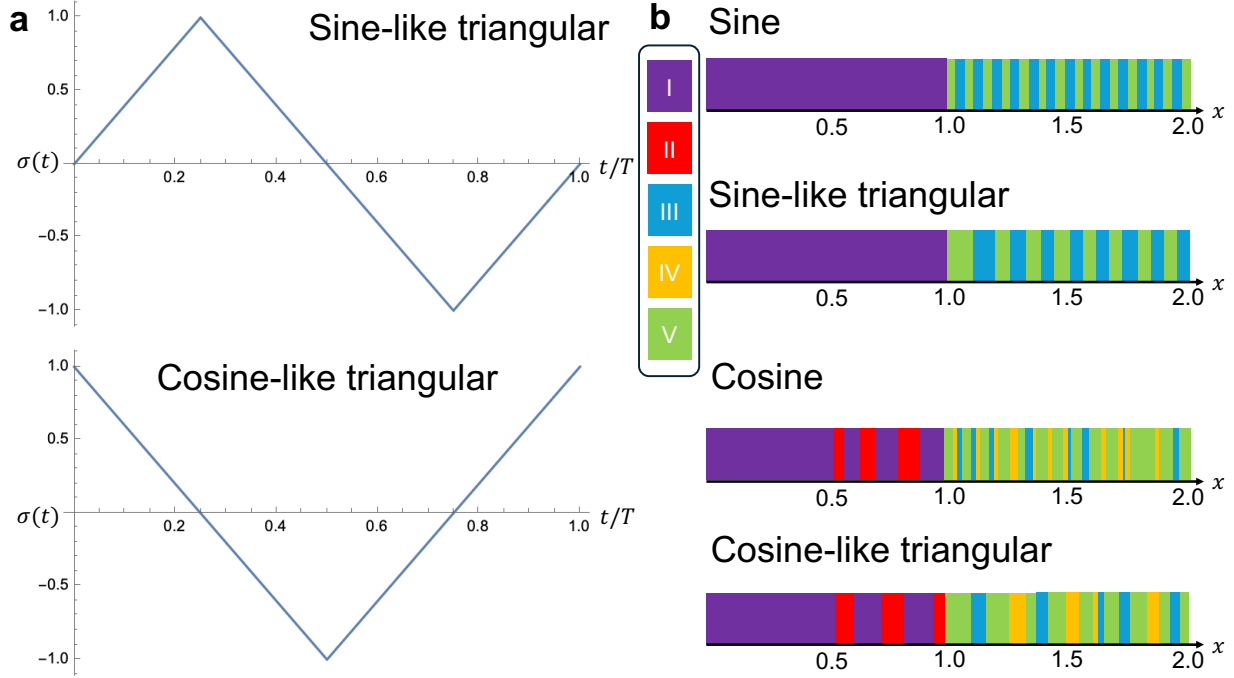

Figure 3: **The magnetization dynamics under triangular waves.** (a) (upper panel) a sine-like triangular wave, (lower panel) a cosine-like triangular wave. (b) The adiabatic mode diagrams at $\omega_{\text{AC}} = 2\pi/240$ and $\alpha = 0.03$ for sine (first), sine-like triangular (second), cosine (third), and cosine-like triangular (fourth) waves. The violet denotes Mode I, the red denotes Mode II, the blue denotes Mode III, the yellow denotes Mode IV, and the green denotes Mode V.

# The time evolution of peaks in Fourier transform

In our manuscript, we describe four types of peaks observed in the Fourier transform of $\phi(t)$. These include: i) the main peak originating from the DC modes denoted as $\omega_1$, ii) a peak corresponding to the AC driven frequency $\omega_2 = \omega_{\text{AC}}$, iii) induced peaks resulting from the frequency difference between i) and ii), given by $\omega_3 = \omega_1 + n(\omega_1 - \omega_2)$, and iv) subpeaks represented by $\omega_4 = |\omega_{1,2,3} \pm 2n\omega_{\text{AC}}|$.

Peaks i) and iii) are associated with the decaying process, whereas peaks ii) and iv) represent the stable state after decay. Figure 4 illustrates this phenomenon. In Fig. 4(a), the Fourier transform is performed from $t = 0$ to $1.26$ $\mu$s, while in Fig. 4(b), it is performed from $t = 0.12474$ to $0.126$ ms. As time progresses, Peaks i and iii diminish due to damping, while Peaks ii and iv remain robust owing to the driven frequency.

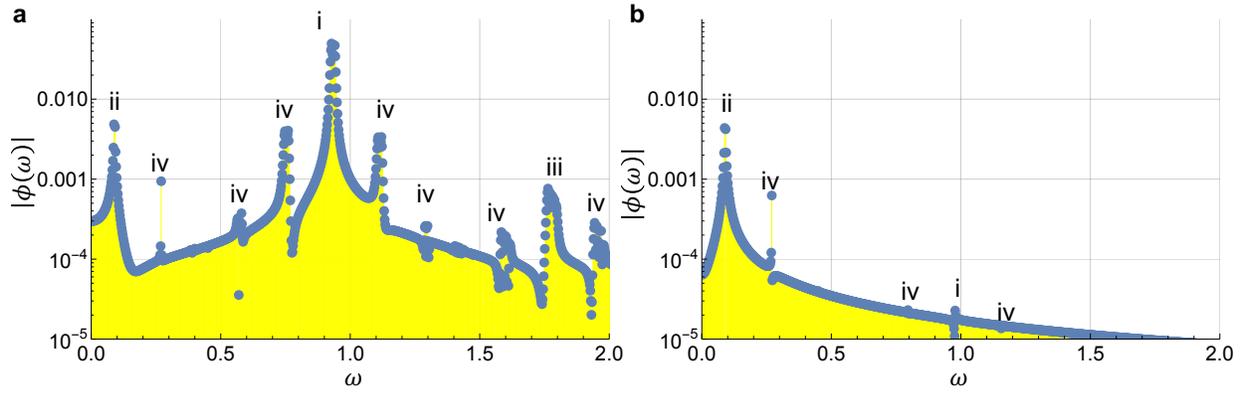

Figure 4: **The time evolution of peaks in Fourier transform of** $\phi(t)$ (a) The Fourier transform of $\phi(t)$ in $t = 0$ to $1.26$ $\mu$s. (b) The Fourier transform of $\phi(t)$ near $t = 0.12474$ to $0.126$ ms.



\bibliography{refer.bib}

\end{document}